\documentclass[%
superscriptaddress,
preprint,
 amsmath,amssymb,
 aps,
prb,
showkeys
]{revtex4-2}

\usepackage{graphicx}
\usepackage{dcolumn}
\usepackage{bm}
\usepackage{afterpage}
\usepackage{color}
\usepackage{ulem}
\usepackage{comment}

\begin{document}


\title{Effect of shear flow on the transverse thermal conductivity of polymer melts}

\author{Kotaro Oda}
 \email{af22q002@guh.u-hyogo.ac.jp}
\author{Shugo Yasuda}%
 \email{yasuda@gsis.u-hyogo.ac.jp}
\affiliation{%
Graduate School of Information Science, University of Hyogo 
}%

\date{\today}

\begin{abstract}
The effect of shear flows on the thermal conductivity of polymer melts is investigated using a reversed nonequilibrium molecular dynamics (RNEMD) method. 
We extended the original RNEMD method to simultaneously produce spatial gradients of temperature and flow velocity in a single direction.
This method enables accurate measurement of the thermal conductivity in the direction transverse to shear flow.

\textcolor{black}{The Weissenberg number defined with the shear rate and the relaxation time of the polymer conformation can uniformly differentiate the occurrence of shear rate dependence of the thermal conductivity across different chain lengths.}
The stress-thermal rule (STR) (i.e., the linear relationship between anisotropic parts of the stress tensor and the thermal conductivity tensor) holds for entangled polymer melts even under shear flows but not for unentangled polymer melts. 
Furthermore, once entanglements form in polymer chains, the stress-thermal coefficient in the STR remains independent of the polymer chain length.

These observations align with the theoretical foundation of the STR, which focuses on energy transmission along the network structure of entangled polymer chains [Brule, Rheol. Acta \textbf{28}, 257 (1989)]. However, under driven shear flows, the stress-thermal coefficient is notably smaller than that measured in the literature for a quasi-quiescent state without external forces. Although the mechanism of the STR in shear flows has yet to be fully elucidated, our study confirmed the validity of the STR in shear flows. This allows us to use the STR as a constitutive equation for computational thermo-fluid dynamics of polymer melts, thus having broad engineering applications.

\end{abstract}

\maketitle
\newpage


\section{\label{sec:level1}INTRODUCTION}
Computer simulations of polymer melts have been actively conducted in the chemical industry in recent years to achieve more efficient processing and product design than traditional experimental approaches. However, accurately predicting the flow behavior and temperature distribution of polymer melts is still challenging because of their complex thermo-rheological nature.
For example, in high-speed polymer processing, a polymer melt has a nonuniform temperature distribution due to significant viscous heating, leading to complicated thermo-rheological behavior.
In fact, in our previous studies\cite{yasuda_yamamoto_2014,yasuda_2019}, a multiscale simulation combining molecular dynamics and computational fluid dynamics demonstrated that polymer melts can exhibit intriguing transitional behaviors in terms of their stress-optical rule.
This behavior arises from the competition between flow deformation and thermal agitation induced by significant viscous heating, both of which influence chain orientation.

Apart from viscous heating, the influence of the thermal conductivity anisotropy, which is an aspect not previously addressed in our previous studies, \textcolor{black}{may also} play a crucial role in the temperature distribution during polymer processing.
The anisotropy of the thermal conductivity of polymer melts has recently been studied by various researchers.

Macroscopically, the thermal conductivity tensor $\boldsymbol{\kappa}$ is defined by
\begin{eqnarray}
  \mathbf{q}=-\boldsymbol{\kappa}\cdot {\nabla T}
  \label{eq:Fouriershear},
\end{eqnarray}
where $\nabla T$ is the temperature gradient and $\mathbf{q}$ is the heat flux vector.
\textcolor{black}{
Throughout this paper, we only consider a weak temperature-gradient regime, in which the thermal conductivity is independent of the temperature gradient.
}
For polymer melts, Brule~\cite{Brule1989} derived a linear relationship between the thermal conductivity tensor and the stress tensor, which is called the stress-thermal rule (STR), based on a network theory for polymer melts assuming that heat is preferentially transmitted along the backbone of a polymer chain.
Specifically, the STR is expressed as follows:
\begin{eqnarray}
  \boldsymbol{\kappa}-\frac13 \mathrm{tr}(\boldsymbol{\kappa})I=\kappa_{eq} C_\mathrm{t} \left(\boldsymbol{\tau}-\frac13\mathrm{tr}(\boldsymbol{\tau})I\right)
  \label{eq:STR},
\end{eqnarray}
where $\boldsymbol{\tau}$ is the stress tensor, $I$ is the identity matrix, $C_\text{t}$ is the stress-thermal coefficient, and $\kappa_\text{eq}$ is the thermal conductivity in the equilibrium state.
Although network theory considers a network structure consisting solely of polymer segments (it ignores interactions between polymer chains that are not connected to the network), experimental and simulation studies
\cite{Choy1978ThermalCO,Choy1999ElasticMA,Simavilla2012,C2SM26788H,Simavilla2014,Broerman1999,nieto2020molecular} have confirmed that the STR indeed holds for the anisotropic parts of the thermal conductivity tensor and the stress tensor for polymer melts under elongational deformation.
An STR has also been confirmed for polymer melts under shear deformation obtained after the cessation of shear flow \cite{Iddir,Schieber2004,Venerus,dai2006}.
However, as far as the authors know, an STR has yet to be confirmed for polymer melts under shear flows.
When an STR is confirmed in a flowing system, it can be used as a constitutive relation for the computational thermo-fluid dynamics of polymer melts, thus having broad engineering applications.

Equation (\ref{eq:STR}) indicates that in the uniform shear flow described as $\nabla \mathbf{v}=\dot\gamma \boldsymbol{\mathrm{i}}\boldsymbol{\mathrm{k}}$ (where $\mathbf{i}$ and $\mathbf{k}$ are the unit vectors in the $x$ and $z$ directions, respectively), the thermal conductivity in the direction transverse to the shear flow, $\kappa_{zz}$, decreases since the normal stress in the direction transverse to the shear flow, $\tau_{zz}$, decreases for polymer melts.
This behavior is completely different from that for simple fluids.
Davis and Evans derived a thermal conductivity tensor for uniform shear flows of Lennard-Jones (LJ) fluids in Ref.~\cite{Evans1993}, which posits that the thermal conductivity transverse to shear flows is written as
\begin{equation}
\kappa^\mathrm{LJ}_{zz}=\kappa_0+c\dot{\gamma}^2, \label{eq:kappa_eff}
\end{equation}
where $\kappa_0$ is the thermal conductivity according to Fourier's law in the quiescent state, $c$ is a small positive constant, and $\dot \gamma$ is the shear rate.
The above equation indicates that for simple fluids, $\kappa^\mathrm{LJ}_{zz}$ is almost constant when the shear rate $\dot \gamma$ is low but increases as $\dot \gamma$ increases when $\dot \gamma$ is sufficiently high.
Thus, the STR could significantly contribute to the distinctive behaviors observed in the thermo-fluid dynamics of polymer melts compared to those of simple fluids.

In this paper, we investigate the STR (i.e., Eq.~(\ref{eq:STR})) of polymer melts under uniform shear flows using molecular dynamics (MD) simulation.
For simple fluids, the thermal conductivity under shear flows has been calculated by various methods.
Davis and Evans utilized a Green-Kubo method to calculate the thermal conductivity in shear flow~\cite{Evans1993}.
Smith employed a wall-driven Couette system with different wall temperatures and determined the thermal conductivity in shear flow by measuring the heat flux and temperature gradient between the walls~\cite{Smith}.
However, in the case of polymer melts, the computational time significantly increases due to the slow dynamics of entangled polymer chains when attempting to calculate the thermal conductivity using the Green-Kubo method.
In the wall-driven approach, the presence of a wall can significantly alter the molecular structure, leading to results that may differ substantially from those of bulk characteristics.

To overcome these issues, we employ the reversed nonequilibrium molecular dynamics (RNEMD) method~\cite{muller1999reversing,muller1997simple}.
This method involves imposing an artificial flux in the system by swapping the momentum (or kinetic energy) of molecules at different locations, which in turn generates a shear (or temperature) gradient as a physical response within the system.
Using the RNEMD method to calculate the thermal conductivity under shear flows has several advantages.
First, it can simultaneously generate a uniform shear flow and a uniform temperature gradient within an MD cell with periodic boundary conditions.
We modified the original RNEMD method to achieve the simultaneous generation of shear flow and a temperature gradient.
This modification allows us to measure the thermal conductivity transverse to the shear flow in the bulk of the polymer melt.
Additionally, the RNEMD method conserves the total momentum and energy of molecules within the MD cell, even when shears and temperature gradients are applied.
This property leads to accurate computations and enables us to significantly reduce the amount of simulation data compared to nonconserved systems such as wall-driven systems.

This paper is organized as follows.
In Sec.~II, we explain our modified RNEMD method for calculating the thermal conductivity in the direction transverse to the shear flow.
Section III provides numerical results for both simple fluids and polymer melts.
In Sec.~III~A, we test the validity of our modified RNEMD method by comparing our numerical results with Eq.~(\ref{eq:kappa_eff}) for LJ fluids.
In Sec.~III~B, the relationships between the thermal conductivity, stress tensor, and bond-orientation tensor are explored \textcolor{black}{for the Kremer-Grest (KG) model polymer melt~\cite{kremer1990}}, and the STR is discussed.
Finally, we present concluding remarks in Sec.~IV.

\section{\label{sec:METHODOLOGY}METHODOLOGY}
\begin{figure}[t]
\centering
  \includegraphics[width=\linewidth]{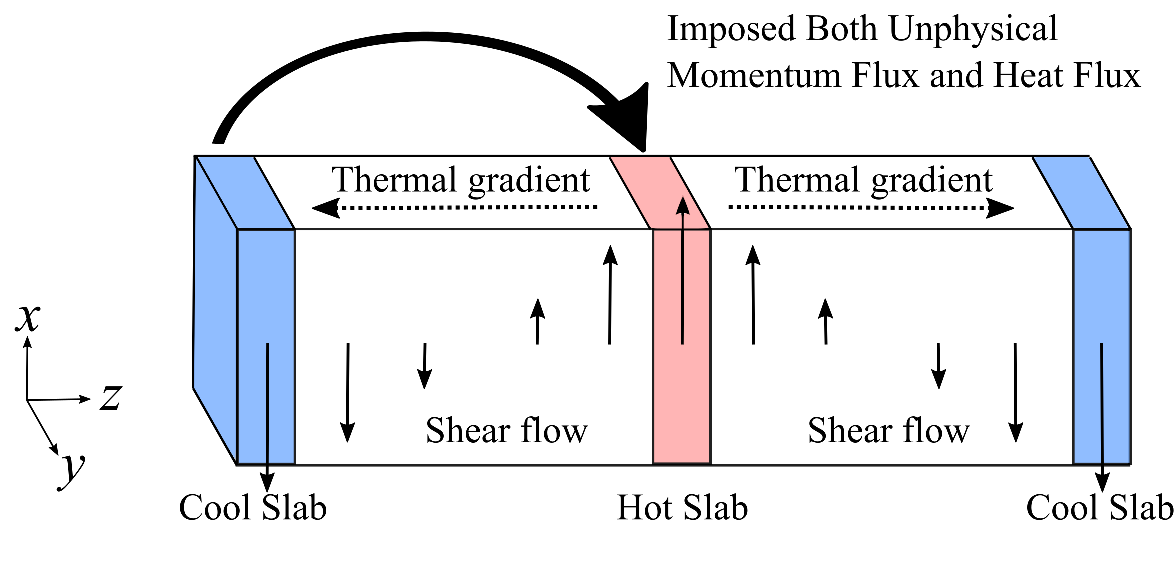}
\caption{
Illustration of our RNEMD method.
We impose both a momentum flux and a heat flux by swapping momentum and kinetic energy between a ``hot'' slab and a ``cool'' slab in the simulation box. The molecular system physically responds to these imposed fluxes by generating both a shear flow and a temperature gradient. }\label{fig:RNEMD}
\end{figure}

We applied the RNEMD method to calculate the thermal conductivity in the transverse direction under shear flow. As shown in Fig.~\ref{fig:RNEMD}, both the heat flux and the $x$-momentum (i.e., the momentum in the $x$-direction) flux are imposed in the $z$ direction in a periodic simulation cell with dimensions $L_x \times L_y \times L_z$.
The simulation cell is divided into slabs along the $z$ axis.
The central slab at $z=L_z/2$ is assigned to the "hot" slab, and the edge slab at $z=0$ and $z=L_z$ is assigned to the "cool" slab.
Additionally, the atoms inside the hot slab are driven in the $+x$ direction, while those inside the cool slab are driven in the $-x$ direction.
The heat flux is generated by swapping the kinetic energy of velocities $y$ and $z$, $\frac{m}{2}({v_y}^2+{v_z}^2)$ ($m$ is the mass of the particle), between the hot and cold slabs at each constant time interval $\Delta t_\text{k}$.
This process involves selecting two atoms: one with the highest kinetic energy in the cool slab and the other with the lowest kinetic energy in the hot slab.
Then, the $y$ and $z$ components of the velocities are swapped within the pair.
Note that in our approach, we do not swap the $x$ component of the velocities in the kinetic energy (KE) swapping.
This differs from the original RNEMD method.
This modification can prevent interference with the shear field driven in the system by $x$-momentum swapping.

The $x$-momentum flux is also generated by swapping the $x$-momentum between the hot and cold slabs at each constant time interval $\Delta t_m$, where we select the atom with the largest $v_x$ in the cold slab and the one with the largest $-v_x$ in the hot slab and then swap the $x$ components of their velocities within the pair.

At the macroscopic level, this swapping process generates \textcolor{black}{heat and momentum fluxes in the $z$-direction, so } spatial gradients of the temperature $T$ and the velocity $v_x$ \textcolor{black}{form} along the $z$ axis in the steady state.
Moreover, we can calibrate the heat flux in the $z$-direction, $q_z$, as
\begin{eqnarray}
  {q}_{z}=\frac{\Delta E}{2tL_xL_y}
  \label{eq:heatflux},
\end{eqnarray}
where $\Delta E$ is the sum of the kinetic energy transferred by these swapping processes during the simulation time $t$ and is calculated by the following equation:
\begin{eqnarray}
  \Delta E=\sum_t\frac{m}{2}(v_\text{h}^2-v_\text{c}^2)
  \label{eq:deltaE},
\end{eqnarray}
where $v_h=\sqrt{v_y^2+v_z^2}$ is the kinetic energy of the particle selected at the hot slab and $v_c=\sqrt{v_y^2+v_z^2}$ is that at the cool slab in the swapping process.
By using the heat flux $q_z$ calibrated as Eq.~(\ref{eq:heatflux}) and the temperature gradient $\langle \frac{\partial T}{\partial z}\rangle$ calibrated from the local temperatures \textcolor{black}{at center and edge slabs, where $\langle\cdot\rangle$ represents the time average}, the thermal conductivity transverse to the shear flow is obtained from Eq.~(\ref{eq:Fouriershear}) as
\begin{eqnarray}
  \kappa_{zz}=-\frac{q_z}{\langle\frac{\partial T}{\partial z}\rangle}.
  \label{eq:Fourier2}
\end{eqnarray}

In this work, MD simulations are performed using the LAMMPS package \cite{LAMMPS}.
The source code of the modified RNEMD method is available from the authors upon request.

\section{\label{sec:sim}SIMULATIONS}
Simulations are performed for the simple LJ fluid and the KG model polymer melt.
In Sec.~IIIA, we test the validity of our method by comparing our RNEMD results with those obtained in a previous study\cite{Smith}.
In Sec.~IIIB, we perform RNEMD simulations for various lengths of polymer chains.
\textcolor{black}{
In addition, the validity of the RNEMD method for calculating the viscosity in the isothermal system and for calculating the thermal conductivity in the quiescent system is tested in Appendix~\ref{sec:SLLOD}.
The appendix investigates the effects of the respective swapping frequencies on the viscosity and thermal conductivity of the LJ fluid and model polymer melts.
The appendix also compares the RNEMD and SLLOD~\cite{sllod1,sllod2} methods for the shear thinning behavior of model polymer melts in the isothermal system.
}

\subsection{Lennard-Jones fluid}\label{sec:simA}
We consider the truncated LJ potential (the so-called Weeks-Chandler-Anderson (WCA) potential~\cite{Week1971}),
\begin{eqnarray}
  U_{\mathrm{LJ}}=
  \begin{cases}
  4\epsilon\left[ \left( \frac{\sigma}{r}\right)^{12} -\left(\frac{\sigma}{r}\right)^{6}\right]+\epsilon&(r<2^{1/6}\sigma)\\
  0 & (r \geq 2^{1/6}\sigma)
  \end{cases}\;
  \label{eq:LJ_WCA}.
\end{eqnarray}
Hereafter, we measure the space and time in units of $\sigma$ and $\sqrt{m\sigma^2/\epsilon}$, respectively.
The temperature $T$ is measured in units of $\epsilon/k_\text{B}$.
The number density $\rho_0=0.85$ and the sizes of the simulation cells $L_x = L_y=10.06$ and $L_z = 40.24$ are fixed.
The rectangular simulation cell is divided into twenty slabs along the $z$ axis.
The equations of motion are integrated using the velocity Verlet algorithm with a time step size of $\Delta t=0.001$.

The system is first equilibrated in the quiescent state at $T=1.25$ for $10^6$ time steps with the Langevin thermostat.
Next, swapping of the kinetic energy is carried out with the time interval $\Delta t_\text{k}=100\Delta t$ for the period $t=10^6\Delta t$ to achieve stationary heat flux.
Finally, the simulation involving heat and momentum fluxes is performed for $10^7$ time steps in a variation of the time interval of the momentum swapping as $\Delta t_m=$100$\Delta t$ -- 1000$\Delta t$.
The time interval of the kinetic energy swapping is fixed as $\Delta t_\text{k}=100$$\Delta t$.
\textcolor{black}{
With $\Delta t_k=100\Delta t$, one can perform accurate computations with low noise in the linear response regime of the temperature gradient for the LJ fluid. 
See Appendix~\ref{sec:SLLOD}.
}

\begin{figure*}[t]
  \includegraphics[width=\linewidth]{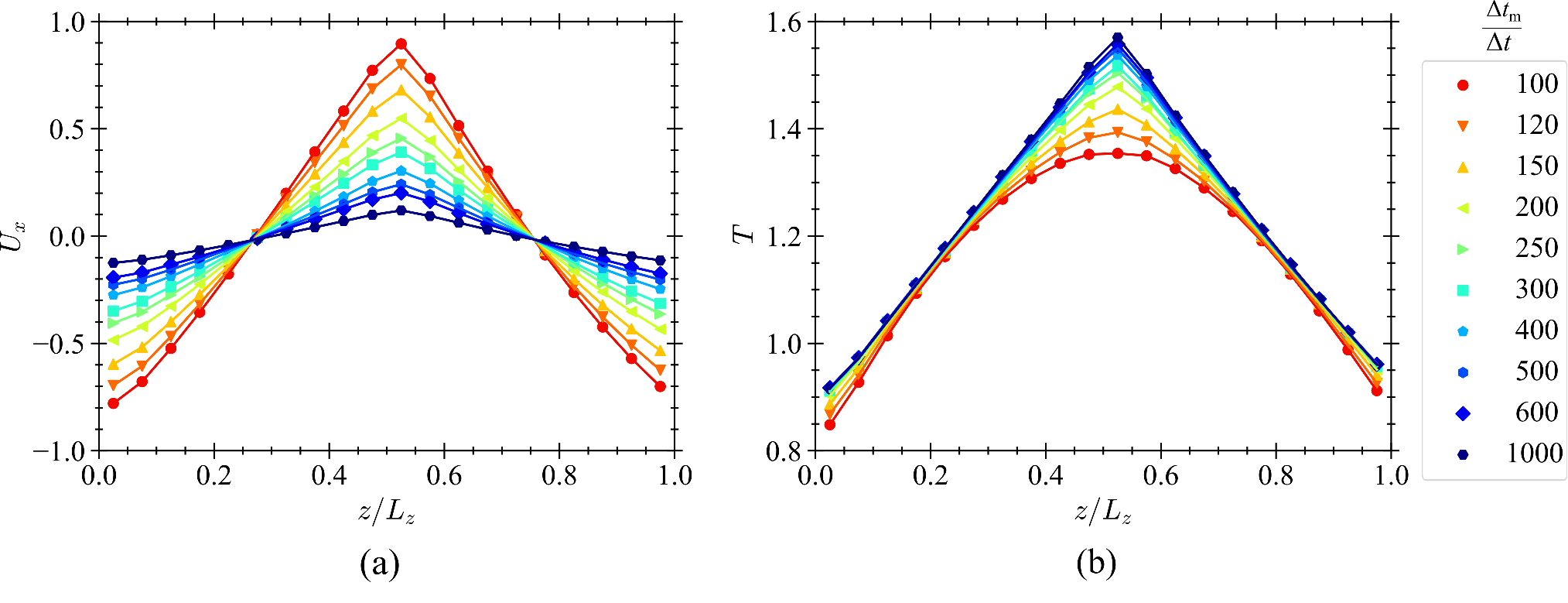}
\caption{
Flow velocity profiles (a) and temperature profiles (b) for the LJ fluid are calculated with different intervals of momentum swapping $\Delta t_\text{m}$, as shown in the legend to the figure.
The interval of the kinetic energy swapping is fixed at $\Delta t_\text{k}$=100$\Delta t$. }\label{fig:LJprofile}
\end{figure*}

\begin{figure}[t]
  \includegraphics[scale=0.8]{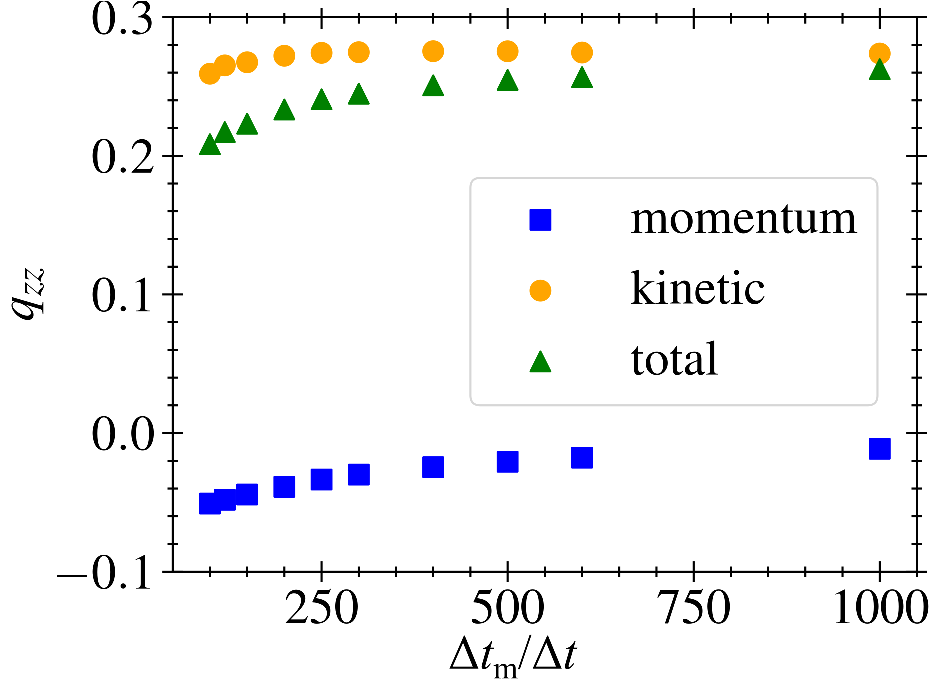}
\caption{
Energy fluxes arising in momentum swapping (blue squares), energy fluxes arising in kinetic energy swapping (orange circles), and total energy fluxes (green triangles) are plotted as functions of the momentum swapping interval.
  Each energy flux is measured in the lower half of the simulation cell ($0<z<L_z/2$).}\label{fig:LJflux}
\end{figure}

\begin{figure}[t]
  \includegraphics[width=0.8\linewidth]{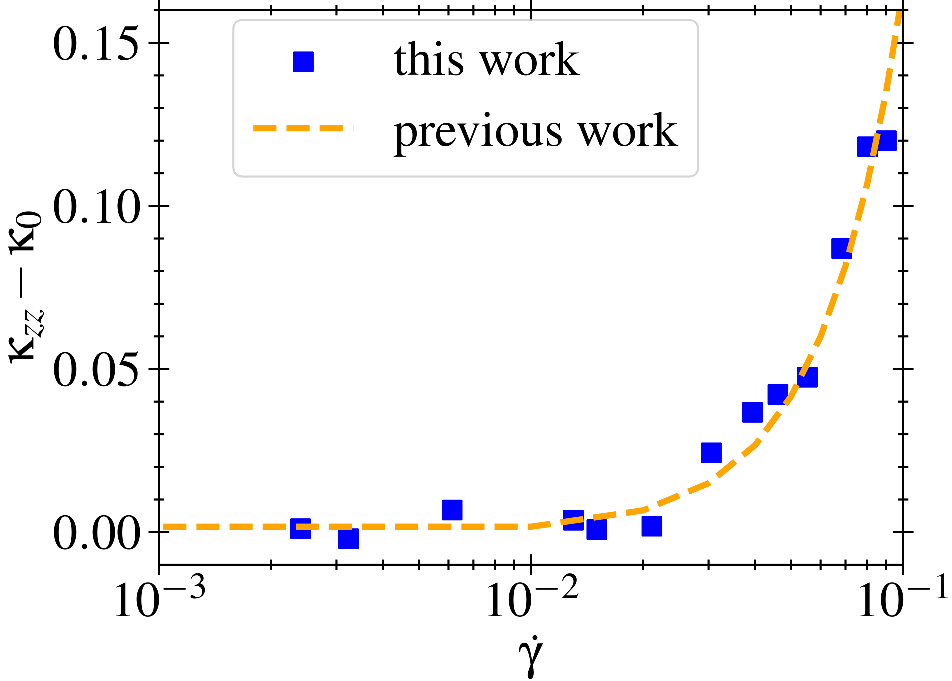}
\caption{
The thermal conductivity of the LJ fluid is plotted as a function of the shear rate $\dot \gamma$.
The blue squares show the results obtained in this study, while the orange dashed line shows the results obtained in Ref. \cite{Smith} (i.e., Eq.~(\ref{eq:kappa_eff})).
}\label{fig:dvdz-th-LJ}
\end{figure}

Figure~\ref{fig:LJprofile} shows the spatial profiles of the flow velocity $U_x$ (in (a)) and the temperature $T$ (in (b)) at different intervals of momentum swapping.
The flow velocity profiles are almost linear irrespective of the swapping interval $\Delta t_m$.
In contrast, the temperature profile becomes rather parabolic as the frequency of momentum swapping $(\Delta t_m)^{-1}$ increases for $\Delta t_m\textcolor{black}{\lesssim} 200$.
This phenomenon was also reported in previous studies\cite{muller1997simple,Tenny}.

It is also observed that the temperature gradient decreases as the frequency of momentum swapping increases.
This behavior is explained in terms of the contribution of momentum swapping to the energy flux.
Swapping atomic velocities $v_x$ between hot and cold slabs may likely induce an 
\textcolor{black}{energy flux in the direction from the hot slab to the cold slab, resulting in a decrease in the temperature gradient}.
In Fig.~\ref{fig:LJflux}, we show the energy fluxes due to the swapping of kinetic energy and those due to the swapping of momentum \textcolor{black}{at different values of $\Delta t_m$} separately.
The energy flux due to kinetic energy swapping is not affected by changing the momentum swapping interval $\Delta t_m$ for $\Delta t_m \gtrsim 200$$\Delta t$.
In contrast, the energy flux due to momentum swapping, which has a negative value, becomes increasingly prominent as the swapping frequency $(\Delta t_m)^{-1}$ increases when $\Delta t_m$ is moderately small (e.g., $\Delta t_m\lesssim 400$$\Delta t$).
\textcolor{black}{As mentioned earlier, momentum swapping may likely induce an energy flux from the hot slab to the cold slab, corresponding to a negative energy flux in the lower half of the simulation cell.}
Thus, the net energy flux decreases as the frequency of momentum swapping increases, causing the temperature gradient to decrease.
\textcolor{black}{The effect of coupling of the energy and momentum fluxes leads to an increase in the thermal conductivity of the LJ fluid as shown in Eq.~(\ref{eq:kappa_eff}).}

Figure~\ref{fig:dvdz-th-LJ} shows \textcolor{black}{the shear rate dependence of the thermal conductivity in the direction transverse to shear flows.}
In this figure, we plot the deviation of $\kappa_{zz}$ from that in the quiescent state, $\kappa_{zz}-\kappa_{0}$, where $\kappa_{0}$ is the thermal conductivity in the quiescent state \textcolor{black}{which is calculated as $\kappa_0=7.76$}, as a function of the shear rate $\dot \gamma$.
Our results are in good agreement with the values obtained in a previous study \cite{Smith} and consistent with the theoretically expected quadratic $\dot\gamma$ dependence \cite{Evans1993} within the wide range of shear rates, even though at small swapping intervals (e.g., $\Delta t_m$=100 and 120), the temperature profiles become rather parabolic.
This result demonstrates that our RNEMD method is valid within the wide range of $\dot{\gamma}$ shown in Fig.~\ref{fig:dvdz-th-LJ}.

Here, we also emphasize that $\kappa_{zz}$ remains almost constant at low shear rates (e.g., $\dot\gamma \lesssim 0.03$) in the LJ fluid 
\textcolor{black}{but increases with $\dot \gamma$ when $\dot\gamma$ is large (e.g., $\dot \gamma>0.03$). 
This high-$\dot\gamma$ regime corresponds to $\Delta t_m/\Delta t < 400$ in Fig.~\ref{fig:LJflux}. Thus, as mentioned earlier, the $\dot \gamma$ dependence of the thermal conductivity of the LJ fluid described as Eq.~(\ref{eq:kappa_eff}) stems from the energy flux accompanied by the exchange of atomistic velocities at the large shear rate.}

\subsection{Polymer melt}\label{sec:simB}
We consider the KG model polymer melt.
The simulation system includes $N_\text{c}$ polymer chains with $N_\text{b}$ beads.
All bead practices interact via the repulsive part of the LJ potential (\ref{eq:LJ_WCA}),
and the consecutive beads on each chain are connected by an anharmonic spring potential
\begin{eqnarray}
  U_{\mathrm{FENE}}=
  -\frac{1}{2}k_cR^{2}_{0} \ln \left[1-\left(\frac{r}{R_0}\right)^{2}\right]
  \label{eq:FENE},
\end{eqnarray}
where $k_{\text{c}}=30$ and $R_0=1.5$ are fixed.
\textcolor{black}{Here, we note that the space, time, and temperature are measured in the same LJ units as in the previous section.}
We carried out simulations of model polymer melts with different chain lengths $N_\text{b}$.
In Table I, the parameter sets used in the simulations are summarized.
The number density $\rho_0=0.85$ and the initial temperature $T_0=1$ are fixed.
\begin{table}[b]
  \caption{\label{tab:tablePara}%
  Parameter values used in the simulations for KG model polymer melts are shown. The first and second columns show the number of beads per molecule and the number of molecules, respectively. 
  The third to fifth columns show the lengths of the simulation cell in the x-direction, y-direction, and z-direction, respectively.
  }
  \begin{ruledtabular}
  \begin{tabular}{ccccc}
  \textrm{$N_\text{b}$}&
  \textrm{$N_c$}&
  \textrm{$L_x$}&
  \textrm{$L_y$}&
  \textrm{$L_z$}\\
  \colrule
  30 & 1600 &24.16&24.16&96.64\\
  50 & 1100 &25.29&25.29&101.16\\
  100 & 600 &26.03&26.03&104.12\\
  200 & 360 &27.66&27.66&110.64\\
    \end{tabular}
  \end{ruledtabular}
\end{table}

Each simulation is performed for $10^8$ time steps with $\Delta t=0.001$.
The time interval of momentum swapping $\Delta t_m$ is between 100 and 600 time steps, while the time interval of kinetic energy swapping $\Delta t_k=200\Delta t$ is fixed.
\textcolor{black}{
With this swapping frequency $\Delta t_k=200\Delta t$, one can perform accurate computations with low nose in the linear response regime of the temperature gradient for model polymer melts.
See Appendix~\ref{sec:SLLOD}.
}

We also note that within this range of $\Delta t_m$ values, the shear rates generated in the RNEMD simulations are, at most, less than 0.005.
\textcolor{black}{Hence, the energy flux accompanied by the exchange of atomistic velocities, which leads to an increase in the thermal conductivity in the large $\dot\gamma$ regime as is observed in Fig.~\ref{fig:dvdz-th-LJ}, is negligible. 
Furthermore, within this range of $\Delta t_m$ values, the shear rates obtained in simulations for each model polymer melt are smaller than the upper bounds of the shear rate $\dot\Gamma$, under which the temperature of the system does not significantly increase due to the viscous heating.
More specifically, $\dot\Gamma$ is estimated as $\dot\Gamma=\sqrt{\kappa_0\Delta T/\eta_0 L_z^2}$ with $\Delta T=1$, where $\kappa_0$ and $\eta_0$ are, respectively, the thermal conductivity and viscosity of the polymer melt in the equilibrated state.
In Table \ref{tab:tablekappa0}, we show $\dot \Gamma$ values for each model polymer melt.}
Thus, in the present study, for polymer melts, we only focus on the small $\dot\gamma$ regime where the higher-order contribution of the shear rate is neglected.

\begin{figure*}[p]
  \includegraphics[width=0.8\linewidth]{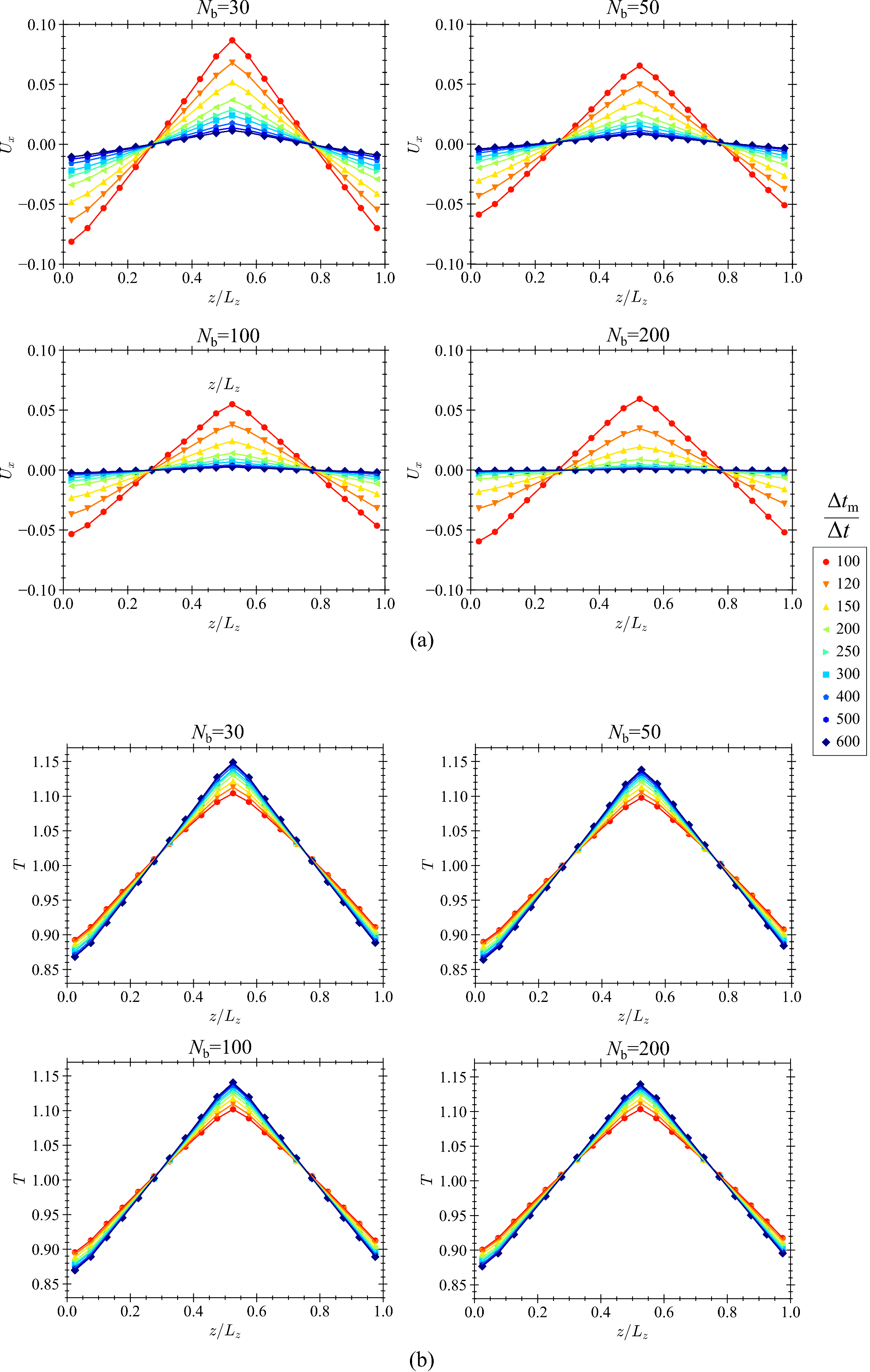}
\caption{
  Flow velocity profiles (a) and temperature profiles (b) calculated for the KG model polymer melt with different momentum swapping intervals $\Delta t_\text{m}$=100$\Delta t$ --- 600$\Delta t$.
The KE swapping interval is fixed at $\Delta t_\text{k}$=200$\Delta t$.}
\label{fig:KGbothprofile}
\end{figure*}

Figure \ref{fig:KGbothprofile} shows the spatial profiles of flow velocity $U_x$ and temperature $T$ for polymer melts with different chain lengths.
Both the velocity and temperature profiles are nearly linear within the parameter range.

\begin{figure}[t]
  \includegraphics[width=1.0\linewidth]{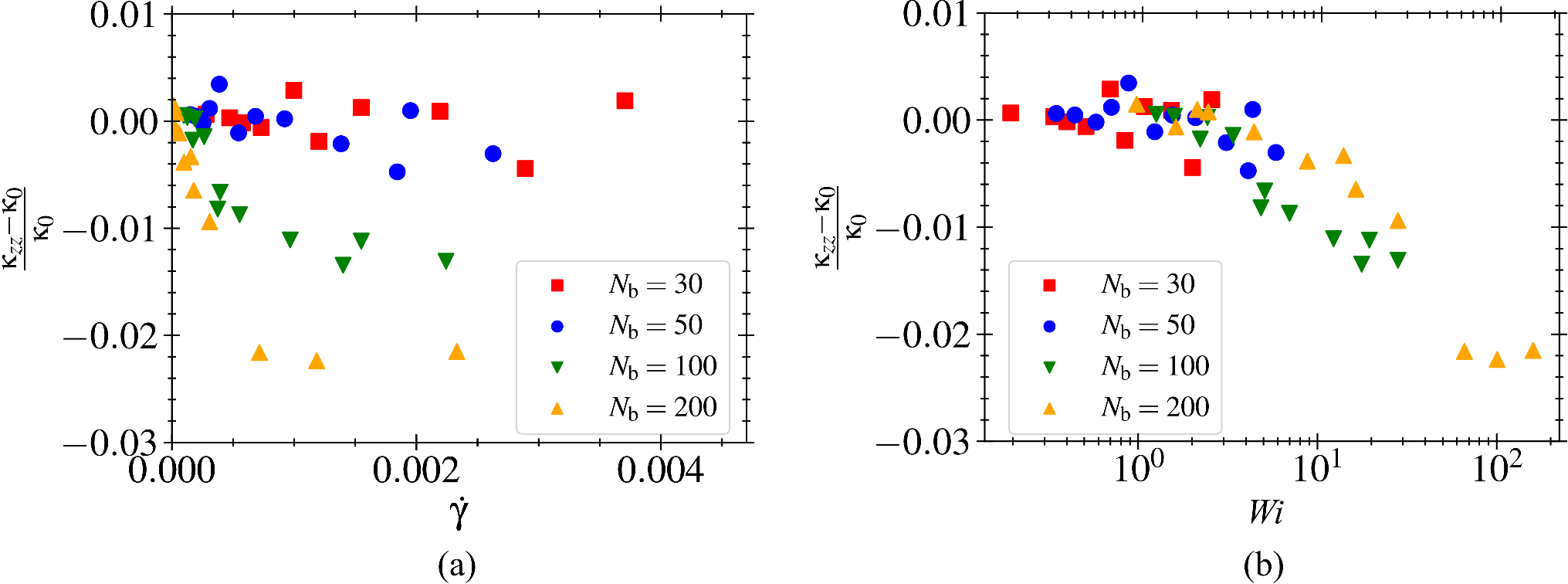}
\caption{\label{fig:dvdz-th-KG} \textcolor{black}{Thermal conductivity plotted as a function of shear rate $\dot \gamma$ (a) and Weissenberg number (b) for the KG model polymer melts.}
The red squares show the results for $N_\text{b}=30$, the blue circles for $N_\text{b}=50$, the green points for $N_\text{b}=100$, and the orange triangles for $N_\text{b}=200$.}
\end{figure}

\begin{table}[b]
  \caption{\label{tab:tablekappa0}%
  Summary of the properties of the KG model polymer melts with different lengths of polymer chains in our simulation system. 
  The first column is the number of beads per molecule. 
  The second and third columns are thermal conductivity and normal stress in the quasi-quiescent states, respectively. \textcolor{black}{The fourth column shows the viscosity in the isothermal state.} 
  \textcolor{black}{The fifth column shows estimated values of the upper bound of the shear rate, under which the viscous heating does not significantly affect the temperature of the system. The sixth column shows the relaxation time of polymer conformation.}}
  
  \begin{ruledtabular}
  \begin{tabular}{cccccc}
\centering
  \textrm{$N_\text{b}$}&
  \textrm{$\kappa_0$}&
  \textrm{$\tau_0$}&
  \textrm{$\eta_0$}&
\textrm{$\dot\Gamma$}&
  \textrm{$t_d$}\\
  
  \colrule
  \centering
  30 &  4.97 & -5.00 &22.1 &4.92$\times 10^{-3}$ &7.40$\times 10^{2}$\\
  \centering 50 &  4.94 & -4.91 & 33.8 &3.83$\times 10^{-3}$&2.38$\times 10^{3}$\\
  \centering 100 &  4.93 & -4.89 & 89.9& 2.26$\times 10^{-3}$&1.32$\times 10^{4}$\\
  \centering 200 &  4.93 & -4.89 & 240&1.30$\times 10^{-3}$&7.39$\times 10^{4}$\\
    \end{tabular}
  \end{ruledtabular}
\end{table}

Figure~\ref{fig:dvdz-th-KG}\textcolor{black}{(a)} shows the thermal conductivity $\kappa_{zz}$ as a function of the shear rate $\dot{\gamma}$.
Here, $\kappa_{0}$ is the thermal conductivity in the quiescent state, and the values of $\kappa_0$ for different chain lengths are summarized in Table \ref{tab:tablekappa0}.
For $N_\text{b}=$30 and $N_\text{b}=$50, the thermal conductivity remains almost unchanged even with an increase in $\dot{\gamma}$.
However, for $N_\text{b}=$100 and $N_\text{b}=$200, the thermal conductivity clearly decreases as the shear rate increases.
{\color{black}
In Fig.~\ref{fig:dvdz-th-KG}(b), the horizontal axis of Fig.~\ref{fig:dvdz-th-KG}(a) has been rescaled by the Weissenberg number $Wi$ for each polymer melt, which is defined as
\begin{equation}\label{eq:Wi}
Wi=t_d\dot\gamma.
\end{equation}
Here, $t_d$ represents the relaxation time of the polymer conformation for each model polymer melt, which is calculated from the time-correlation function of the end-to-end vector of the polymer chain (See Appendix~\ref{sec:APP}).
The values of $t_d$ at different chain lengths are shown in Table \ref{tab:tablekappa0}.

Interestingly, the results across different chain lengths exhibit similar behavior.
The Weissenberg number can differentiate the occurrence of shear rate dependence of the thermal conductivity; the thermal conductivity is almost constant when $Wi\lesssim 3$ but decreases with $\dot\gamma$ when $Wi\gtrsim 3$.
This indicates that the shear rate affects the thermal conductivity when the memory of polymer conformation is preserved during the time scale of shear flow, i.e., $t_d \gg \dot \gamma^{-1}$.
Thus, relaxation of polymer conformation is a key factor for the shear rate dependence of the thermal conductivity.

We also note that the shear rate dependence of the thermal conductivity is not evident for short chains with $N_b=30$ and 50.
This is because the relaxation time of the polymer conformation $t_d$ for short chains is much smaller than that for long chains. 
Consequently, $Wi$ is not sufficiently large for the shear rate dependence of thermal conductivity to become evident.
Indeed, for short chains, $Wi <3 $ always holds in the small $\dot\gamma$ regime considered in this section, where both the effect of coupling of the heat and momentum fluxes (as observed in the LJ fluid) and the effect of viscous heating can be ignored (see also Table~\ref{tab:tablekappa0}).
}

\begin{figure*}[t]
  \includegraphics[width=\linewidth]{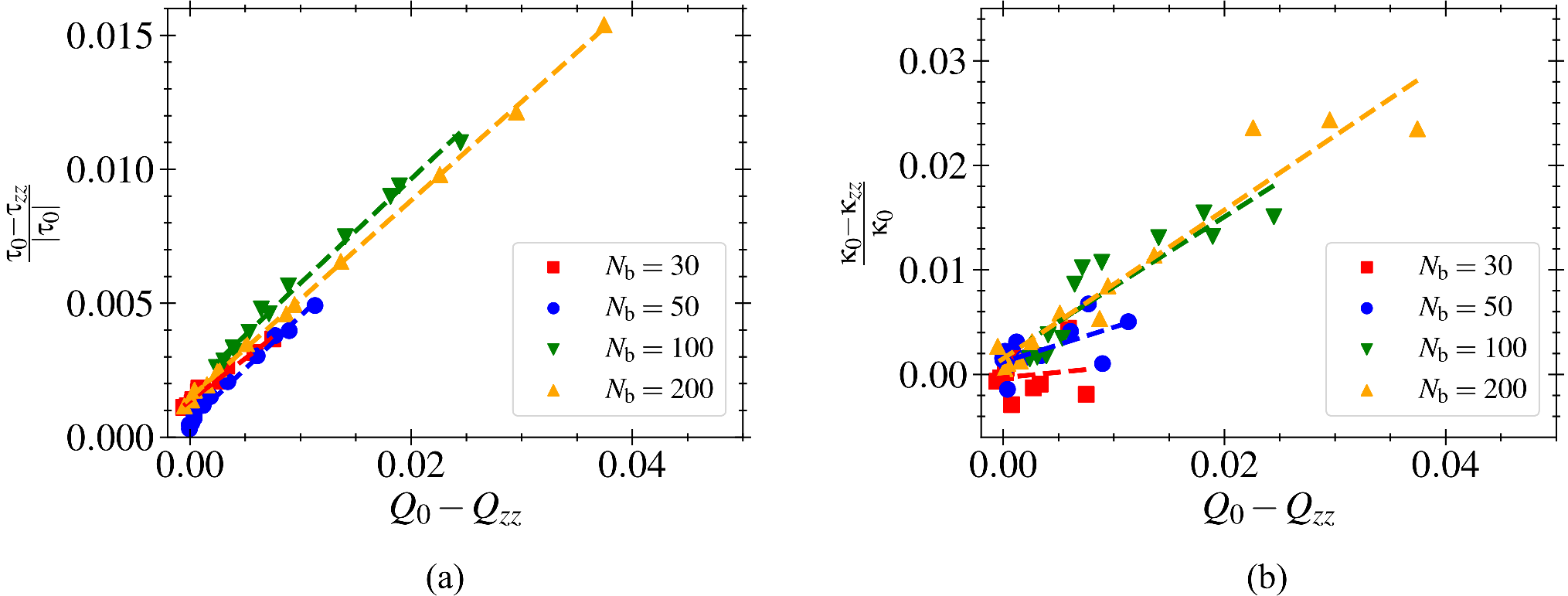}
\caption{
The deviations in the normal stress (a) and thermal conductivity (b) from those in the quiescent state are plotted as functions of the deviation of the bond orientation from that in the quiescent state.
The red squares show the results for the KG model polymer melt with $N_\text{b}=30$, the blue circles show those for $N_\text{b}=50$, the green downward triangles show those for $N_\text{b}=100$, and the orange upward triangles show those for $N_\text{b}=200$.
}\label{fig:bond-stress-th}
\end{figure*}

Figure~\ref{fig:bond-stress-th} shows the relationship between the normal stress difference and bond orientation [in (a)] and the relationship between the thermal conductivity and bond orientation [in (b)].
In each figure, the dashed lines represent the linear fits of the simulation results for different chain lengths.
Here, the bond-orientation tensor $Q_{ij}$ ($i,j=\{x,y,z\}$) is defined as
\begin{eqnarray}
  Q_{ij}=\frac{1}{N_\text{c}}\sum_\text{chains}\frac{1}{N_\text{b}-1}\sum_{k=1}^{N_\text{b}-1}\frac{b_{i}^{k}}{b_\text{min}}\frac{b_{j}^{k}}{b_\text{min}}
  \label{eq:bori},
\end{eqnarray}
where ${b_{i}^{k}}$ for $1 \leq k \leq N_\text{p}-1$ is the component of the bond vector between consecutive beads in the same chain and $b_\text{min}$ is the distance at which the total potential $U_\text{LJ}(r) +U_\text{FENE}(r)$ is minimized (i.e., $b_\text{min} \simeq 0.97$).
On the horizontal axis, $Q_0$ is the value in the quiescent state (i.e., $Q_0\simeq 0.33$).
The stress tensor $\tau_{ij}$ ($i,j=\{x,y,z\}$) is defined as
\begin{eqnarray}
  \tau_{ij}= -\frac{1}{V}\sum_{k }\left[m(v_{i}^k-U_{i}^k)(v_{j}^k-U_{j}^k)
  +\frac{1}{2}\sum_{l\neq k}F_{i}^{kl}r_{j}^{kl}\right]
  \label{eq:stress},
\end{eqnarray}
where $v_{i}^k$ and $U_{i}^k$ denote the atomic velocity of the $k$th bead and the macroscopic flow velocity at the position of the $k$th bead, respectively.
The force acting on the $k$th bead by the $l$th bead is written as $F_{i}^{kl}$.
Here, $\tau_0$ is the normal stress in the quiescent state, and the values of $\tau_0$ for polymer melts with different chain lengths are summarized in Table \ref{tab:tablekappa0}.

The normal stress difference is linearly proportional to the bond orientation for all the polymer melts with different numbers of beads $N_\text{b}$. Thus, we obtain the well-known stress-optical rule given by
\begin{eqnarray}
  \frac{\tau_{0}-\tau_{zz}}{|\tau_{0}|}= C_\text{s}(Q_0-Q_{zz}),
  \label{eq:SOR}
\end{eqnarray}
where $C_\text{s}$ is the stress-optical coefficient.
The values of $C_\text{s}$ are calculated as 0.32, 0.39, 0.39, and 0.37 for $N_\text{b}=30$ $N_\text{b}=50$, $N_\text{b}=100$ and $N_\text{b}=200$, respectively.
Our simulation results show that the linear stress-optical rule robustly holds irrespective of the length of the polymer chain even if a small thermal gradient exists.

In contrast, for the relationship between the thermal conductivity and the bond orientation, the linear dependence of the thermal conductivity on the bond orientation is not clearly observed for short chains $N_\text{b}=30$
and $N_\text{b}=50$, whereas it appears that the linear relationship holds for long chains $N_\text{b}$=100 and $N_\text{b}=$200, as expressed by
\begin{eqnarray}
  \frac{\kappa_{0}-\kappa_{zz}}{\kappa_{0}}= C_\text{b}(Q_0-Q_{zz}),
  \label{eq:bond-kappa}
\end{eqnarray}
where $C_\text{b}$ is the thermal-optical coefficient.
Furthermore, once the linear relationship holds at $N_\text{b}\gtrsim 100$, $C_\text{b}$ is not affected by the beard number $N_\textrm{b}$.
The values of $C_\text{b}$ are calculated as 0.67 and 0.71 for $N_\text{b}=100$ and $N_\text{b}=200$, respectively.

Notably, the present KG model polymer melt represents unentangled polymers at $N_\text{b}$=30 and 50 but entangled polymers at $N_\text{b}$=100 and $N_\text{b}=$200 (see Appendix).
This indicates that the linear relationship between the thermal conductivity and the bond-orientation tensor holds only for entangled polymers and does not hold for unentangled polymers.

This behavior is intuitively explained as follows:
Microscopically, the thermal conductivity of a polymer melt is determined mainly via two contributions: the interaction of inconsecutive bead particles (as in the LJ fluid) and the transmission of molecular vibrations in consecutive bead particles along polymer chains.
The results of the LJ fluid in Sec.~IIIA indicate that the former contribution does not affect the thermal conductivity in the small \textcolor{black}{$\dot\gamma$ regime (i.e., $\dot\gamma <\dot\Gamma$)} considered in this section.
Thus, it is thought that the deviation of the thermal conductivity from that in the quiescent state, $\kappa_0-\kappa_{zz}$, stems from the latter contribution.
Once entanglement is generated between polymer chains, \textcolor{black}{the characteristics of the polymer conformation are preserved for a long relaxation time $t_d$, which may be much larger than the flow time scale $\dot\gamma^{-1}$.
Thus, as is assumed in the derivation of the STR [Eq.~(\ref{eq:STR})]~\cite{Brule1989}, the direction of the transmission of molecular vibrations is characterized by the bond orientation tensor.}
Indeed, as predicted by the STR, the deviation of the thermal conductivity $\kappa_0-\kappa_{zz}$ linearly depends on that of the bond orientation tensor $Q_0-Q_{zz}$ when entanglement is generated between polymer chains.

\begin{figure*}[ht]
  \includegraphics[width=0.7\linewidth]{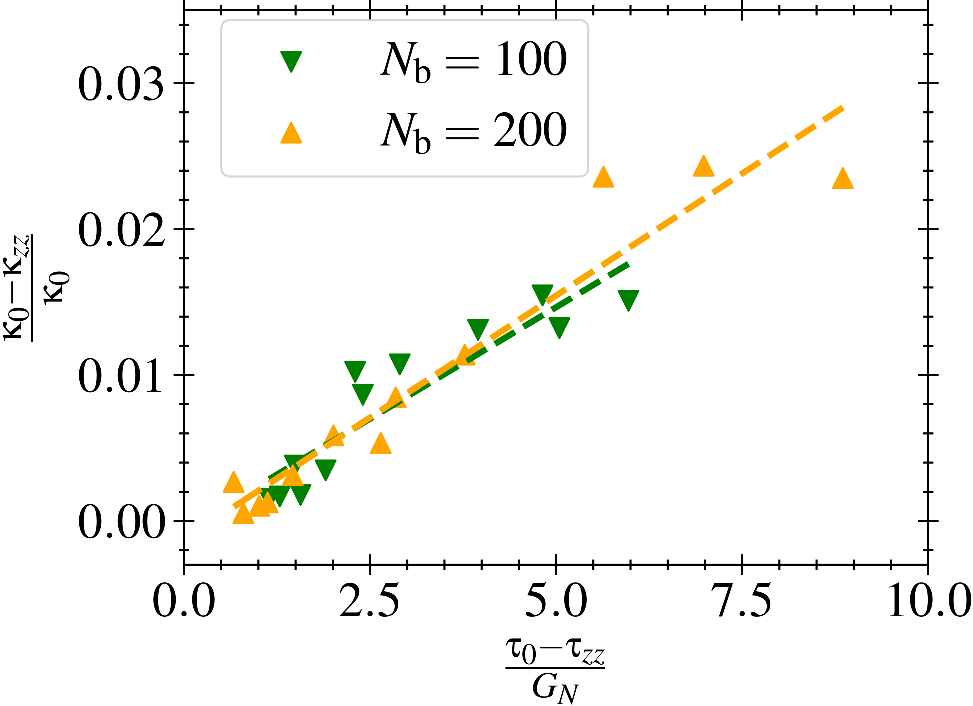}
\caption{The deviation of the thermal conductivity from that in the quiescent state is plotted as a function of the deviation of the normal stress from that in the quiescent state for entangled model polymer melts.
The green downward triangles show the results for $N_\text{b}=100$, and the orange upper triangles show the results for $N_\text{b}=200$.}
  \label{fig:STR}
\end{figure*}

Finally, Figure~\ref{fig:STR} shows the relation between the thermal conductivity $\kappa_{0}-\kappa_{zz}$ and the stress difference $\tau_0-\tau_{zz}$ expressed by
\begin{eqnarray}
  \frac{\kappa_{0}-\kappa_{zz}}{\kappa_{0}}= C_\text{t}G_\text{N}\frac{\tau_0-\tau_{zz}}{G_\text{N}}
  \label{eq:STR0-zz},
\end{eqnarray}
where $G_\text{N}$ is the plateau modulus and is estimated to be $8.0 \times 10^{-3}$ for our simulation model (see Appendix).
The above equation is slightly different from the STR described by Eq.~(\ref{eq:STR}).
In Eq.~(\ref{eq:STR0-zz}), the trace of the thermal conductivity tensor $\mathrm{tr}(\boldsymbol{\kappa})$ in Eq.~(\ref{eq:STR}) is replaced with $\kappa_0$.

For entangled polymer chains $N_\text{b}=$100 and $N_\text{b}=$200, the linear relationship (\ref{eq:STR0-zz}) holds, and their coefficients $C_\text{t} G_\text{N}$ are almost the same.
This behavior is consistent with the experimental results obtained in Refs.~\cite{2013Gupta,C2SM26788H,Simavilla2012,Simavilla2014,Venerus2001}, where the STR is measured under deformation in quasi-quiescent states after the cessation of external forces.
However, in our results, the value of the coefficient is an order of magnitude smaller than the experimental result $C_\text{t}G_\text{N}\simeq 0.04$ obtained in the references.
In our results, the values of the coefficient are calculated as $C_\text{t}G_\text{N}=3.1\times10^{-3}$ and $3.3\times 10^{-3}$ for $N_\text{b}=100$ and $N_\text{b}=200$, respectively.
The reason for this discrepancy has not been explained.

\afterpage{\clearpage}

\section{\label{sec:CON}CONCLUDING REMARKS}
We investigated the STR of polymer melts in the transverse direction under shear flows by using molecular dynamics simulations.
We extended the original RNEMD method to simultaneously generate spatial gradients of flow velocity and temperature in a single direction.
This method enables accurate measurement of the thermal conductivity in the direction transverse to shear flow.
The modified RNEMD method was first verified through comparisons with theoretical and numerical results, given in the references, on the nonlinear dependence of the thermal conductivity on the shear rate for LJ fluid.
Subsequently, this approach was applied to KG model polymer melts.

\textcolor{black}{
For the LJ fluid, the thermal conductivity is almost constant concerning the variation in the shear rate except in the large shear-rate regime.
However, in the large shear-rate regime (say $\dot\gamma >0.03$ in the present study), the thermal conductivity transverse to shear flow increases with the shear rate.
This occurs because exchanges of atomic velocities, which generate a strong shear flow, in a temperature gradient system likely induce a heat flux from a high-temperature region to a low-temperature region, resulting in an increase in the thermal conductivity in the direction transverse to the shear flow.
Thus, the increase in the transverse thermal conductivity of the LJ fluid with the shear rate, which is observed in the large $\dot\gamma$ regime, is due to the coupling of heat and momentum fluxes.
}

\textcolor{black}{
On the other hand, for model polymer melts, the shear rate affects the thermal conductivity even at a small $\dot\gamma$ when the polymer chain is sufficiently long.
The $\dot\gamma$ dependence appears when the Weissenberg number, which is defined as $Wi=t_d\dot\gamma$, where $t_d$ is the relaxation time of the polymer conformation, is sufficiently large (say $Wi>3$ in the present model polymer melts).
This indicates that the $\dot \gamma$ dependence of the thermal conductivity for polymer melts stems from a completely different mechanism than that for the LJ fluid, and the relaxation of the polymer conformation is an important key factor for the STR.
}

Our main finding is that the STR holds for entangled polymer melts even under \textcolor{black}{weak} shear flows, but it does not hold for unentangled polymer melts.
This behavior completely differs from the stress-optical rule (which uniformly holds for polymer melts irrespective of chain length).
The fact that the occurrence of an STR is linked to the occurrence of polymer chain entanglement aligns with the derivation of an STR in the network theory of polymer melts that considers energy transmission along the network structure of entangled polymer chains.

Furthermore, our numerical results demonstrated that once entanglement occurs in polymer chains, the stress-thermal coefficient is not affected by the chain length.
This observation is consistent with the experimental results of anisotropic thermal conductivity measured in a quasi-quiescent state \textcolor{black}{after cessation of} shear or elongational flows.
However, our numerical results indicate that the stress-thermal coefficient $C_t G_N$ in Eq.~(\ref{eq:STR0-zz}) is one order of magnitude smaller in shear flows than in the quasi-quiescent state.
The mechanism of the decrease in the stress-thermal coefficient in shear flows has yet to be explained.
However, our observation seems reasonable because the energy transmission along the network of polymer chains should be hindered by successive breakages of the network structure during shear flow.

\textcolor{black}{
In the present study, we only investigate the small $\dot\gamma$ regime for polymer melts, in which the effect of coupling of heat and momentum fluxes, which leads to an increase in the thermal conductivity transverse to shear flow, and the effect of viscous heating are ignored.
In this small $\dot\gamma$ regime, the thermal conductivity of the polymer melt is characterized by the orientation of polymer chains when the Weissenberg number is sufficiently large.
Since the orientation of polymer chains is dominant in the flow direction, the transverse thermal conductivity decreases with the shear rate.
However, in the large $\dot \gamma$ regime, as mentioned above, there is a mechanism that may induce an increase in the transverse thermal conductivity with the shear rate.
Thus, a competitive mechanism seems to exist in the effect of shear rate on the transverse thermal conductivity of polymer melts.
The STR in the large $\dot \gamma$ regime is an interesting open question.
}

Apart from the physical aspects of the STR, this study fundamentally contributes to the computer simulation of the thermo-fluid dynamics of polymer melts.
\textcolor{black}{Although the anisotropy of the thermal conductivity may only slightly affect the temperature distribution in weak shear flows,}
by utilizing either the STR or the thermal-optical relation as the constitutive relation for the energy flux
instead of the conventional Fourier’s law with an isotropic thermal conductivity, 
one can compute the coupled dynamics of flow behavior and temperature distribution \textcolor{black}{at least} more accurately.
\textcolor{black}{
Although it has yet to be confirmed in either experiments or simulations, if the STR holds even in high Weissenberg number flows (e.g., $Wi>100$), then the anisotropy of the thermal conductivity may play a crucial role} in broad engineering applications, for example, polymer processing in chemical engineering and lubrication systems in mechanical engineering.
In a future study, the distinct behaviors produced by the STR in thermo-fluid dynamics will be addressed.

Finally, in the present study, we considered only the STR in the transverse direction to the shear flow.
However, the original STR is given in tensor form.
Thus, to confirm the STR, we need to investigate the longitudinal direction of the shear flow.
This will also be addressed in our future study.

\appendix
\section{\label{sec:APP}Relaxation functions of unentangled and entangled polymer melts}
In Appendix A, we investigate the differences in the relaxation behavior between the model polymer melts composed of short chains (i.e., $N_\text{b}=30$ and 50) and those composed of long chains (i.e., $N_\text{b}=100$ and 200) in the equilibrium state.
Furthermore, we present the method for estimating $G_\text{N}$ in Eq.~(\ref{eq:STR0-zz}).
The relaxation dynamics of short-chain models with $N_\text{b}<N_\text{e}$, where $N_\text{e}$ represents the average bead number between consecutive entanglements, can be described by Rouse dynamics, while those of long-chain models with $N_\text{b}>N_\text{e}$ can be described by reptation dynamics~\cite{Doi}.
The value of $N_\text{e}$ is typically in the range of 60 to 100~\cite{kremer1990,Lee2009, M.Pütz_2000, Yamamoto2004}.

\begin{figure*}[ht]
  \includegraphics[width=0.55\linewidth]{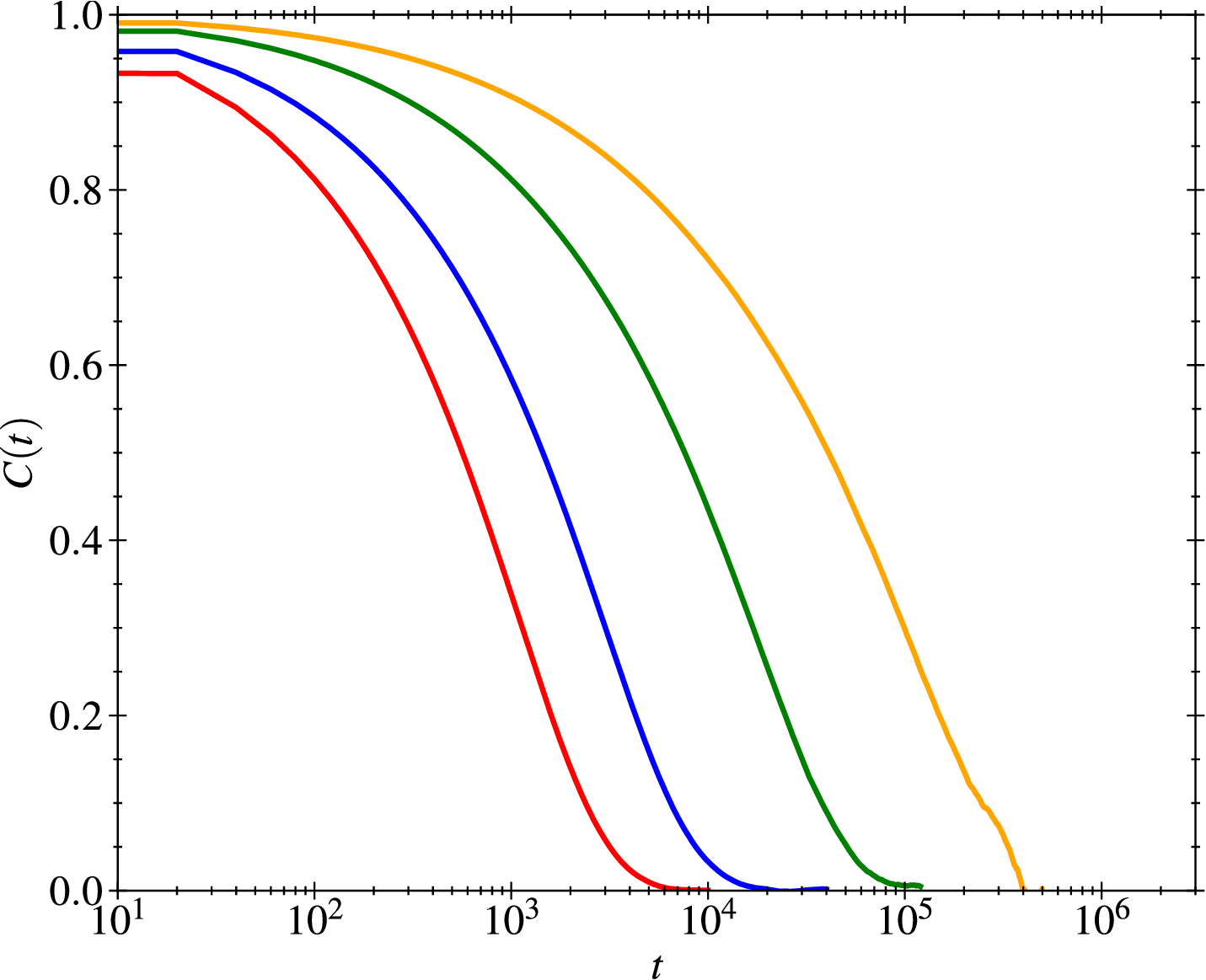} \caption{Normalized time-correlation function of the end-to-end vectors $C(t)$ for the KG model polymer melts with $N_\text{b}=30$ (red line), $N_\text{b}=50$ (blue line), $N_\text{b}=100$ (green line) and $N_\text{b}=200$ (orange line). }
  \label{fig:Ct-RELAX}
\end{figure*}

\begin{figure}[ht]
  \includegraphics[width=0.55\linewidth]{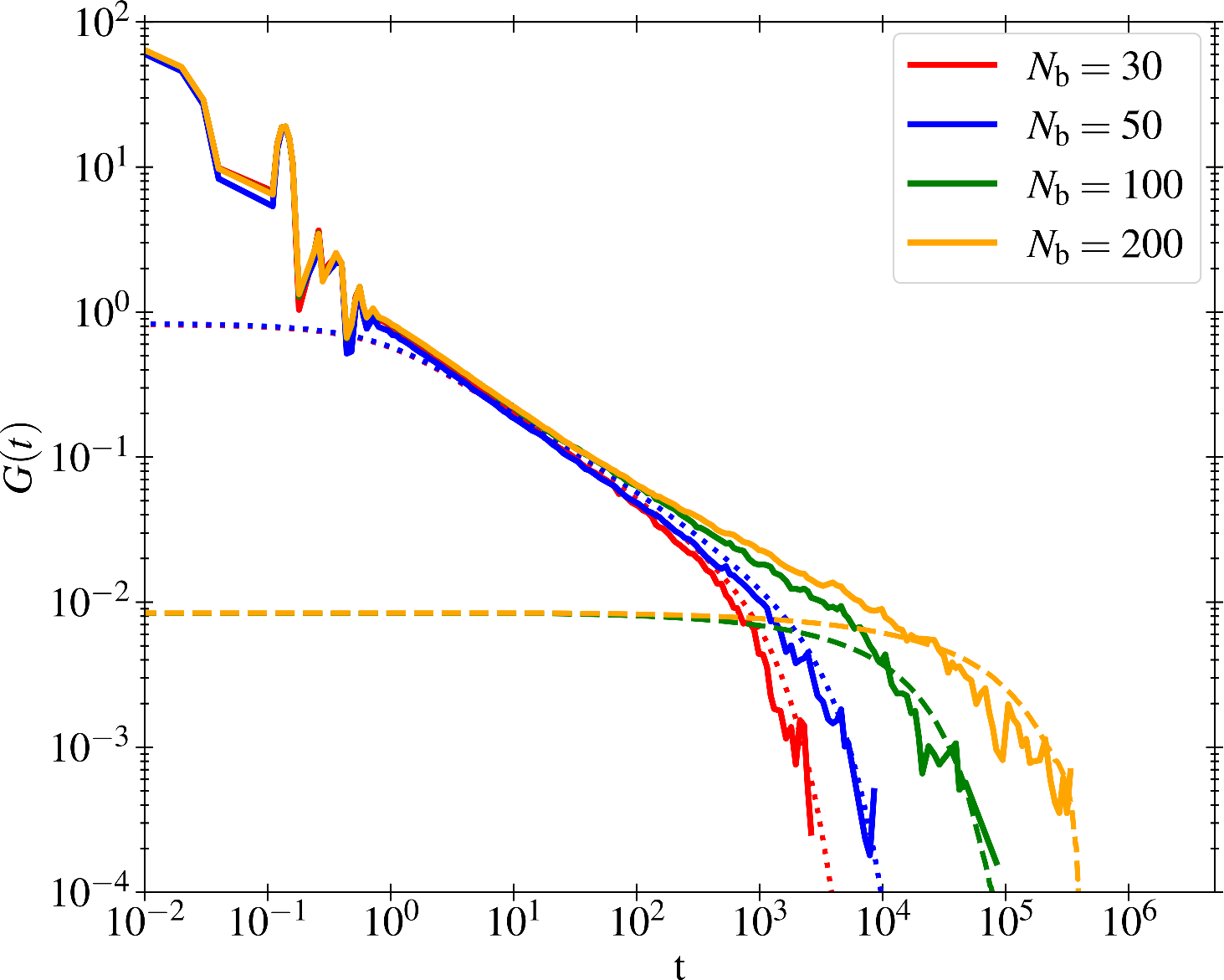}
\caption{Stress autocorrelation function $G(t)$ for the KG model polymer melts with $N_\text{b}=30$ (red solid line), $N_\text{b}=50$ (blue solid line), $N_\text{b}=100$ (green solid line) and $N_\text{b}=200$ (orange solid line). The Rouse relaxation function $G_\text{Rouse}(t)$ in Eq.~(\ref{eq:Rouse}) for $N_\text{b}=30$ (red dotted line) and $N_\text{b}=50$ (blue dotted line) and the reptation relaxation function $G_\text{rep}(t)$ in Eq.~(\ref{eq:Rep}) for $N_\text{b}=100$ (green dashed line) and $N_\text{b}=200$ (orange dashed line) are also plotted.}
  \label{fig:Gt-RELAX}
\end{figure}

Figure \ref{fig:Ct-RELAX} shows the normalized time-correlation function of the end-to-end vector of the polymer chain $C(t)$ calculated as
\begin{eqnarray}
  C(t)=  \frac{\langle \mathbf{P}(t+t_0) \cdot \mathbf{P}(t_0)\rangle}{\langle \ |\mathbf{P}(t_0)|^2\rangle} 
  \label{eq:vec},
\end{eqnarray}
where $\mathbf{P}(t)$ represents the end-to-end vector of each polymer chain.
Figure \ref{fig:Gt-RELAX} shows the stress relaxation function $G(t)$ calculated as
\begin{eqnarray}
  G(t)= \frac{V}{3k_\text{B}T}(\langle \tau_{xy}(t+t_0)\tau_{xy}(t_0)\rangle+
  \langle \tau_{xz}(t+t_0)\tau_{xz}(t_0)\rangle+\langle \tau_{yz}(t+t_0)\tau_{yz}(t_0)\rangle)
  \label{eq:stress-autocorrelation}.
\end{eqnarray}
According to $C(t)$, both the Rouse dynamics and the reptation dynamics predict the same simple functional form in terms of the relaxation time $t_d$ as follows:

\begin{eqnarray}
  C(t)= \sum_{\text{odd} \: p} \frac{8}{\pi^2p^2}\exp(-p^2t/t_d)
  \label{eq:ct_th},
\end{eqnarray}
where the summation is over odd $p$ and $1 \leq p \leq N_{\text{b}}-1$. Since the first term $p=1$ in the summation is dominant for any $t$, this function decays nearly exponentially, and $\tau_\text{d}$ may be estimated from $C(t_d) \simeq \exp(-1)$.
We obtained the relaxation times $t_d$ for $N_\text{b}=30,50,100$ and $200$ as
$t_d$=7.40$\times 10^{2}$, 2.38$\times 10^{3}$, 1.32$\times 10^{4}$, and 7.39$\times 10^{4}$.

\textcolor{black}{
Theoretically, in reptation dynamics, the stress relaxation function $G_{\text{rep}}(t)$ is given by \cite{Doi,Fetters,Larson}
\begin{eqnarray}
  G_\text{rep}(t)= \frac{4}{5}\frac{\rho k_{B}T}{N_\text{e}}C(t) 
  \label{eq:Rep}.
\end{eqnarray}
Here, the models with $N_\text{b}=100$ and $N_\text{b}=200$ are assumed to be represented by reptation dynamics.  
In this paper, $N_\text{e}$ is estimated as $N_\text{e}=85$ for both $N_\text{b}=100$ and $N_\text{b}=200$ by fitting Eq.~(\ref{eq:Rep}) to our numerical results in Fig.~\ref{fig:Gt-RELAX} at $t=t_d$.}

On the other hand, in Rouse dynamics, the stress relaxation function $G_\text{Rouse}(t)$  is expressed as

\begin{eqnarray}
  G_\text{Rouse}(t)= \frac{\rho k_{B}T}{N_\text{b}}\sum_{p=1}^{N_\text{b}-1}
\exp(-p^2t/\textcolor{black}{t_{R}}) 
  \label{eq:Rouse},
\end{eqnarray}
where $t_\text{R}$ is the Rouse relaxation time.
The models with $N_\text{b}=30$ and $N_\text{b}=50$ should be represented by Rouse dynamics because $N_\text{b} \ll N_\text{e}$, which means that $t_\text{R}$ should be equal to $t_d$ for $N_\text{b}=30$ and $N_\text{b}=50$. 

In Figure \ref{fig:Gt-RELAX}, we also compare our simulation results and the theoretical formulas with the fitting values obtained above.
Our numerical results $G(t)$ for $N_\text{b}=30$ and $N_\text{b}=50$ exhibit Rouse dynamics, while those for $N_\text{b}=100$ and $N_\text{b}=200$ closely match the reptation dynamics.

The plateau modulus $G_\text{N}$ is determined by the following equation:
\begin{eqnarray}
  G_\text{N}= \frac{4}{5}\frac{\rho k_{B}T}{N_\text{e}}
  \label{eq:Plateau}.
\end{eqnarray}
Thus, the value of $G_\text{N}$ is estimated as $8.0 \times 10^{-3}$ for both $N_\text{b}=100$ and $N_\text{b}=200$.

\afterpage{\clearpage}

\newpage
{\color{black}
\section{\label{sec:SLLOD} Further validation of the RNEMD method}

This appendix presents further validations of the RNEMD method.
We evaluate the thermal conductivity $\kappa_0$ and the viscosity $\eta$ separately as a function of their respective swapping frequencies $\Delta t_k$ and $\Delta t_m$ for both the LJ fluid and model polymer melts to investigate their dependency on these parameters.
Furthermore, we compare the viscosity of model polymer melts using a more conventional algorithm, SLLOD \cite{sllod1,sllod2} to verify the expected shear thinning behavior.
Note that the thermal conductivity is calculated in the quiescent state while the viscosity is calculated in the isothermal system.

In Fig. \ref{fig:lj_tk_tm}(a), we show the dependency of the thermal conductivity on the energy swapping frequency $\Delta t_k$ for the LJ fluid. 
The thermal conductivity is almost unchanged for $\Delta t_k/\Delta t \ge 100$. 
However, at high swapping frequencies $\Delta t_k/\Delta t < 100$, it exhibits an evident dependence on $\Delta t_k$.
The former (i.e, $\Delta t_k/\Delta t \ge 100$) corresponds to the linear response regime in which the thermal conductivity is independent of the thermal gradient, while the latter (i.e., $\Delta t_k/\Delta t<100)$ corresponds to the nonlinear response regime where the thermal conductivity depends on the thermal gradient.
In Sec.~\ref{sec:simA}, we fix $\Delta t_k=100\Delta t$ for the LJ fluid because $\Delta t_k=100\Delta t$ is the optimal value to perform an accurate computation with low noise in the linear response regime.

Figure \ref{fig:lj_tk_tm}(b) shows the dependence of the viscosity on the momentum swapping frequency $\Delta t_m$ for the LJ fluid.
Unlike the thermal conductivity $\kappa_0$, the viscosity $\eta$ remains nearly unchanged for all values of $\Delta t_m$. 
This indicates that in the isothermal system, the viscosity of the LJ fluid is nearly independent of the shear rate in the range of swapping frequency $\Delta t_m$ examined in this paper.

The results of the thermal conductivity and viscosity for model polymer melts are shown in Fig. \ref{fig:kg_tk_tm}(a) and (b), respectively.
The thermal conductivity of model polymer melts does not depend on the chain length $N_\text{b}$.
It is almost unchanged for $\Delta t_k/\Delta t \ge 200$ while, for $\Delta t_k/\Delta t < 200$, it exhibits an evident dependence on $\Delta t_k$.
This behavior is similar to that observed in the LJ fluid.
In Sec.~\ref{sec:simB}, we fix the energy swapping frequency for model polymer melts at the optimal value $\Delta t_k=200\Delta t$, which enables accurate computations with low noise in the linear response regime.

Figure \ref{fig:kg_tk_tm}(b) shows the dependence of the viscosity of model polymer melts on the momentum swapping frequency $\Delta t_m$. 
The viscosity of model polymer melts clearly depends on $\Delta t_m$, and the dependence becomes increasingly strong as the chain length $N_b$ increases.
This behavior corresponds to shear thinning of polymer melts.

To verify the shear thinning behavior obtained by the RNEMD method, we compare the RNEMD results with those obtained by the SLLOD algorithm in Fig. \ref{fig:RNEMDvsSLLOD}.
The RNEMD results show good agreements with those obtained by the SLLOD algorithm for all values of $N_\text{b}$ over the wide range of shear rates. 
Therefore, we conclude that the RNEMD method can correctly reproduce the expected shear thinning behavior of model polymer melts.

\begin{figure}[ht]
  \includegraphics[width=1\linewidth]{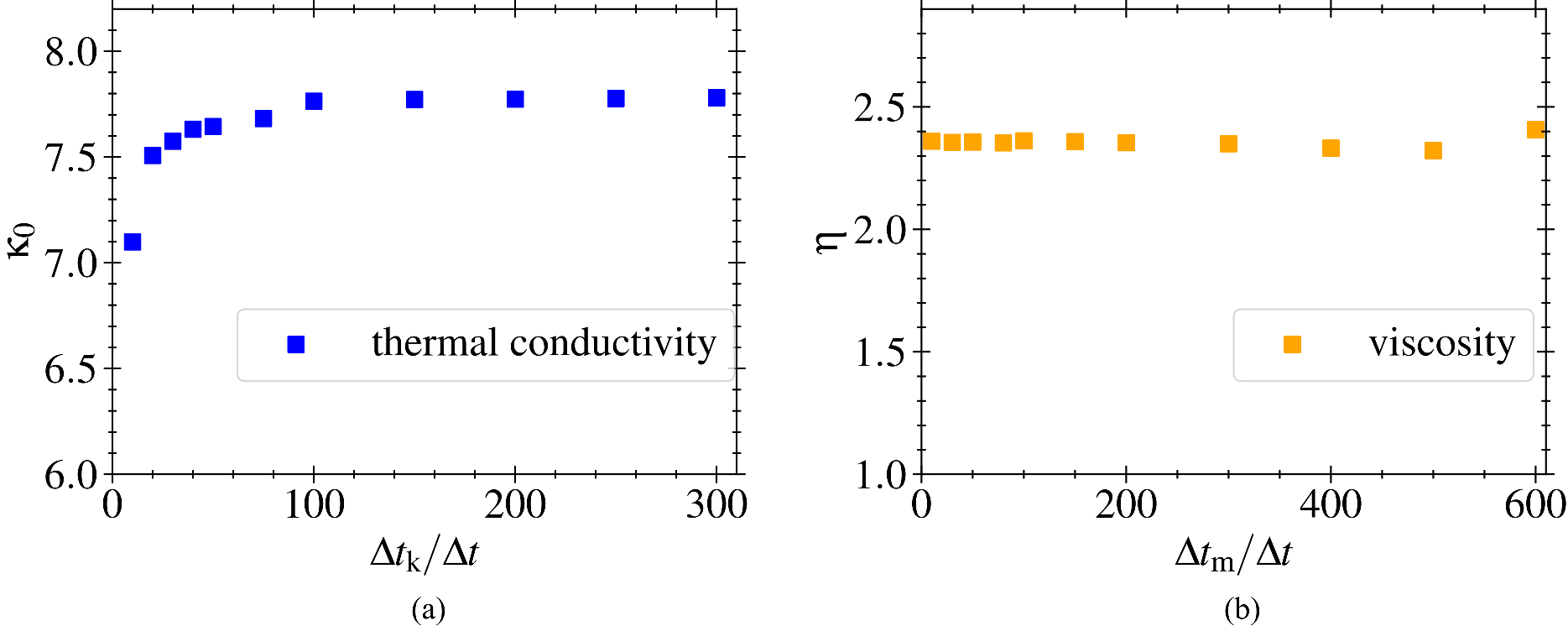}
\caption{\textcolor{black}{Thermal conductivity (a) and viscosity (b) as a function of their respective swapping frequencies $\Delta t_k$ and $\Delta t_m$ in the LJ fluid.}}
  \label{fig:lj_tk_tm}
\end{figure}

\begin{figure}[ht]
  \includegraphics[width=1\linewidth]{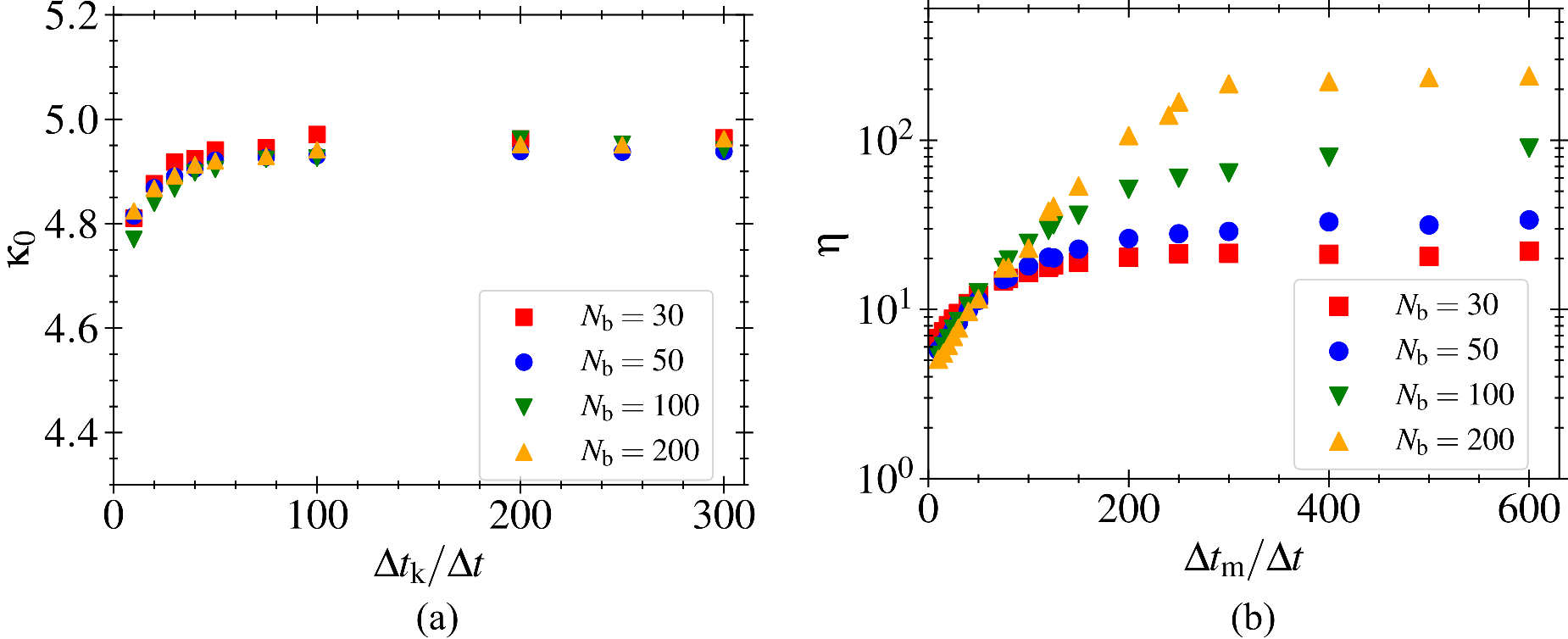}
\caption{\textcolor{black}{Thermal conductivity (a) and viscosity (b) as a function of their respective swapping frequencies $\Delta t_k$ and $\Delta t_m$ for model polymer melts with $N_\text{b}=30$ (red), $N_\text{b}=50$ (blue), $N_\text{b}=100$ (green) and $N_\text{b}=200$ (orange).}}
  \label{fig:kg_tk_tm}
\end{figure}

\begin{figure}[ht]
  \includegraphics[width=0.8\linewidth]{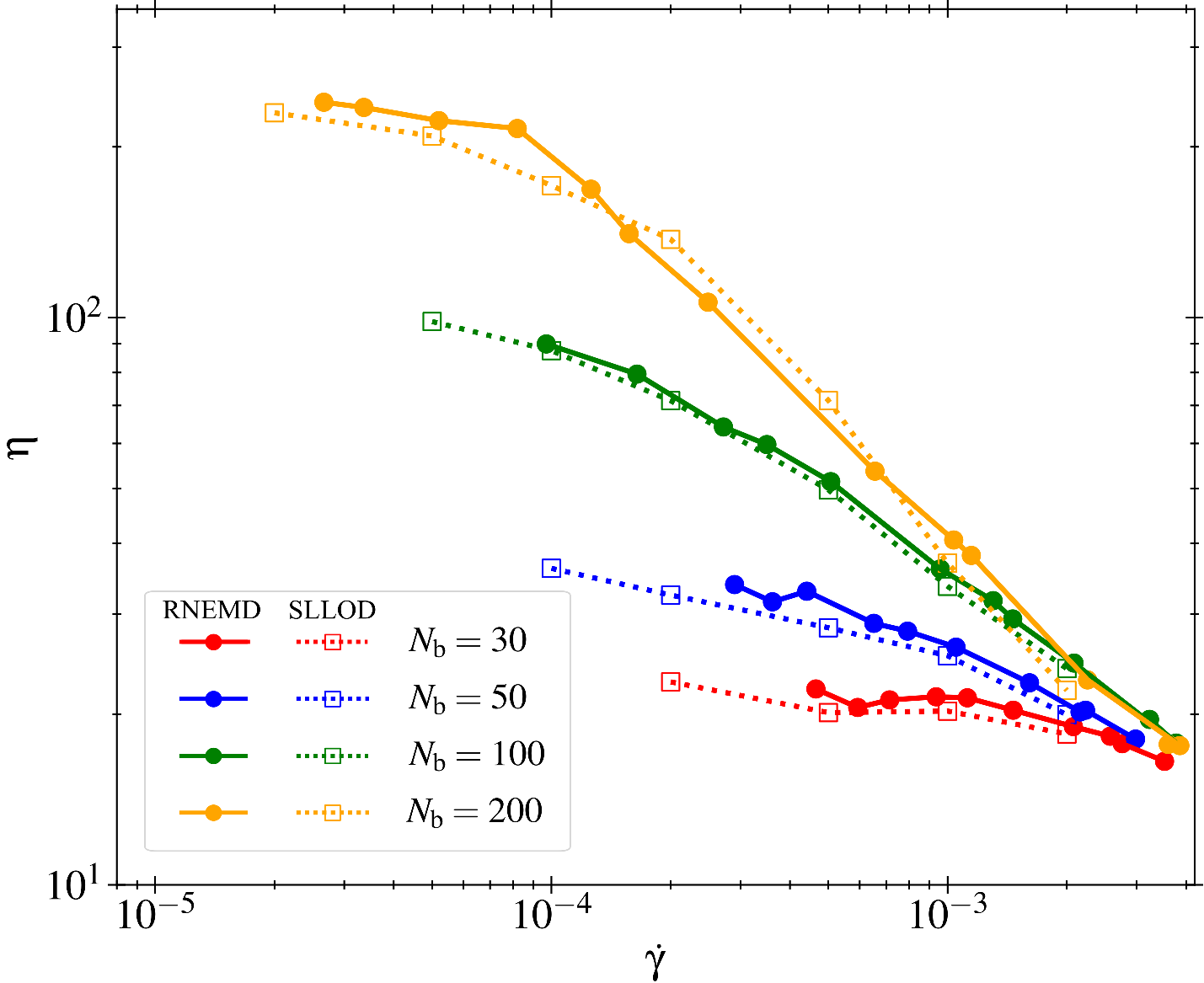}
\caption{\textcolor{black}{Comparison of the shear thinning behaviors of model polymer melts obtained by the RNEMD method (sold lines) and the SLLOD method (dotted lines).
Results of different chain lengths with $N_\text{b}=30$ (red), $N_\text{b}=50$ (blue), $N_\text{b}=100$ (green) and $N_\text{b}=200$ (orange) are shown.}}
  \label{fig:RNEMDvsSLLOD}
\end{figure}
}

\clearpage

\begin{acknowledgements}
The computations in this work were performed using the supercomputers of ACCMS, Kyoto University and the Center for Cooperative Work on Data Science and Computational Science, University of Hyogo.
This work was supported by JSPS KAKENHI Grant Number 23H03350.
\end{acknowledgements}

\nocite{*}

\afterpage{\clearpage}

\bibliography{RNEMD_KG}

\end{document}